\documentclass{article}

\usepackage{amsmath}
\usepackage{amsfonts}
\usepackage{amssymb}

\newcommand{\mc}[1]{\mathcal{#1}}
\newcommand{\mb}[1]{\mathbb{#1}}
\newcommand{\nada}{\varnothing}

\newcommand{\ch}[1]{#1^{\#}}

\title{A Cooperative Proof of Work Scheme for Distributed Consensus
  Protocols}
\author{Wouter Kuijper}

\begin{document}

\maketitle

\section{Introduction}

We propose a refinement to the well known, and widely used,
proof-of-work scheme of \emph{zeroing a cryptographic
  hash}~\cite{zohar2015bitcoin,decker2013information}. Our refinement
is interesting because it allows multiple autonomous users to
\emph{cooperate} on the proof-of-work for their own transactions in
order to bring about consensus on the order of said transactions.

Our scheme is specifically designed to allow for easy cooperation. In
contrast, on the unmodified version of the problem of zeroing a
cryptographic hash cooperation is possible only through the additional
external machinery of \emph{mining pools}. Moreover, these solutions
suffer from problems where participants are actually incentivized
\emph{not} to cooperate~\cite{lewenberg2015bitcoin}.

An inherently cooperative scheme would instead allow us to replace
\emph{transaction fees} (which are paid \emph{to} transaction miners)
by \emph{transaction taxes} (which are paid \emph{by} transaction
miners). This could, in turn, replace competition among miners (which
can have a potentially inflationary effect on power usage) with
frugal, minimal effort, cooperative strategies among users (which can
have a potentially mitigating effect on power usage).

Other potential benefits of our refinement include: 
\begin{itemize}
\item increased defense against discrimination of (certain groups of)
  users by transaction miners,
\item increased throughput of the system as a whole in handling
  transactions (because of reduced competition among miners),
\item increased deterrence against DoS attacks (which will become
  costly due to the transaction tax).
\end{itemize}

In the remainder of this paper we describe the problem of distributed
consensus, we explain the role of proof-of-work in solving the
distributed consensus problem and we formalize our proposed
cooperative proof-of-work scheme.

\section{Consensus}

The problem of distributed consensus arises in the context of
peer-to-peer networks. So let us assume we have a number of peers that
can communicate on a network. To make this more concrete we may say,
for example, that peers are using a \emph{gossiping protocol} to
disseminate information to each other~\cite{demers1987epidemic}.

In addition assume that there is a requirement to keep a
\emph{distributed ledger} containing transactions (of some unspecified
nature). All peers can fully autonomously decide to join the public
ledger and request transactions to be appended to it, but, at the same
time, all peers must agree on what exactly are the contents of the
ledger at all times. Finally, and crucially, these requirements are to
be fulfilled without relying on any form of central authority. In
particular this means that using some a priori leader election
protocol is not an acceptable solution.

One of the first problems that we see is that messages sent by one
peer take some time to propagate through the network and become common
knowledge among all peers.

First note that, as long as the period in-between transactions is,
statistically, \emph{sufficiently higher} than the time it takes a
message to become common knowledge, there is not a big problem in
achieving consensus. All that peers need to do, in that case, is to
cache all messages they have received and then wait until they have
not received anything for at least two full message propagation
periods.
When this happens they can be reasonably sure that other peers have
seen all the same messages, and, crucially, other peers have seen the
pause (the \emph{full stop} if you will) in the overall traffic. This
shared observation can be leveraged to obtain a de-facto consensus on
a set of transactions that can then be safely sorted by some canonical
criterion (like a cryptographic hash) and appended, in that order, to
the distributed ledger.

Unfortunately, we cannot assume that the frequency with which peers
are sending each other transactions is anywhere near low enough for
this scheme to work. 

At this point, proof-of-work schemes can be helpful. Proof-of-work
schemes generally consist of mathematical puzzles (i.e.: finding an
input to a cryptographic hash that starts with a required number of
zeros). In particular, these puzzles are based on problems that we
assume can only be solved by brute-force search. 

Because of the latter property, proof-of-work constitutes a guarantee
that the party providing it has expended a lot of effort in obtaining
it. But, more interestingly for us, it also places an upper-bound on
the frequency with which a peer (with only finite hashing power) can
produce such proof-of-work. Moreover, by adjusting the level of work
required, we can calibrate this upper-bound arbitrarily low as our
network requires for achieving distributed consensus.

As such, proof-of-work can be a tool to force the frequency of
transactions down to a level where these transactions may serve as a
the basis for a de-facto distributed consensus.

\section{Cooperative Proof of Work}

In this section we formalize our cooperative proof-of-work scheme that
addresses the problem of bringing down the frequency of message
streams (in order to achieve distributed consensus).

\subsection{Blocks}

Let $\mc{P}$ be a set if public keys that allows to identify
participants uniquely and check their cryptographic signatures.

Let $h(\cdot)$ be some, suitably strong, cryptographic hash
function. We assume that $h$ is defined for all finite, atomic objects
that we will introduce in the remainder. In addition we assume
$h(\cdot)$ is defined for finite products by hashing the concatenation
of the (self delimiting) canonical binary representation of the
individual elements.

We now give a recursive definition of the set of \emph{blocks}, which
can be seen as the containing objects for the possible transactions on
our distributed ledger.

First we define the set $B_0$ of all \emph{basic blocks}.
To this end we first define a set of \emph{basic entries} $E$ that
participants can request to be entered into the distributed ledger. We
will not consider the internal structure or semantics of the set of
basic entries here (since this is very application specific).

We now define a \emph{basic block} as a a signed pair $(r, e)_p$ for
some public key $p \in \mc{P}$, some basic entry $e \in E$ and some
nonce $r \in \mb{N}$. The interpretation is that a basic block
represents a request by user $p$ to enter entry $e$ into the
distributed ledger, $r$ is a nonce that will be used as a
proof-of-work as explained below.

For every basic block $b \in B_0$ we define the characteristic hash
$\ch{b} = h(r, h(e, p))$.
We let $z \in \mb{N}$ be our \emph{cryptographic hash zero threshold
  parameter}. We require $z$ to be strictly less than the length of
our cryptographic hash values.
We now say a basic block $b \in B_0$ is \emph{proved} iff
the initial $z$ binary digits of $\ch{b}$ are zero. 
With $B_0^{\#}$ we denote the set of all \emph{proved basic blocks}.

Under this definition, proved basic blocks must be \emph{mined}.  This
means that a user needs to do \emph{brute force search} in order to
find a suitable nonce $r$ such that the latter condition holds. It
follows that $z$ must be sufficiently low to allow a single user to
mine basic blocks with adequate frequency and at acceptable power
levels. We will leave open what ``adequate'' and ``acceptable'' mean
precisely in this context (since this is very application specific).

Next we define the set $B_n$ of \emph{compound blocks of level $n$} in
terms of proved blocks of level $n-1$ as follows.
We let $d \in \mb{N}$ be our \emph{cryptographic hash nesting depth
  parameter}. We require $d \ge 2$, in order to avoid some degenerate
cases in the definitions that follow.
We now define a \emph{compound block of level $n$} as a signed
sequence of proved sub-blocks $(b_1, b_2, \dots, b_d)_p$ for some
public key $p \in \mc{P}$ and some $b_1, b_2 \dots, b_d \in
B_{n-1}^{\#}$.

For every compound block $(b_1, b_2, \dots, b_d)_p \in B_n$ we
recursively define the characteristic hash using $d$ nested
applications of our cryptographic hash function $\ch{b} = h(\ch{b_1},
h(\ch{b_2}, h(\dots h(\ch{b_d}, p))))$.
We now say a compound block $b \in B_n$ is \emph{proved} iff the
initial $z+n$ binary digits of $\ch{b}$ are zero (or all digits in
case $z+n$ is greater than the length of the hash).
With $B_n^{\#}$ we denote the set of all \emph{proved compound blocks
  of level $n$}.

Under this definition, proved compound blocks must be \emph{mined}, as
before. In this case, the mining entails that a user needs to do brute
force search in order to find a suitable \emph{ordered sequence} of
sub-blocks such that the latter condition holds. By our definition
each increase in level makes the mining of compound blocks precisely
\emph{twice as hard} (as long as we do not exceed the length of our
cryptographic hash values).

Note that mining a compound block is different from mining a basic
block in that it necessarily involves mixing sub-blocks together. As a
nice side-effect, this gives miners that do not discriminate between
sub-blocks a competitive advantage over miners that do. This effect
becomes more pronounced as we increase the $d$ hashing depth
parameter. The latter observation is the reason we expect our scheme
to offer a better defense against discrimination of (groups of) users
by other (groups of) users.

We wrap up our definition of blocks by defining $\mc{B} = \bigcup_{n
  \in \mb{N}} B_n^{\#}$ as the set of \emph{all proved blocks}. In
addition we define $\mc{B}_{\ge m} = \bigcup_{n \in \mb{N}.n \ge m}
B_n^{\#}$ as the set of \emph{all proved blocks of level $n$ or
  higher}.

\subsection{Streams}

We define a \emph{block stream} as a function $A : \mb{R}^+ \to
2^{\mc{B}}$ mapping instances in time (as positive real numbers)
to sets of proved basic/compound blocks arriving at that time instance
(from the perspective of a single peer). We require $A(0) = \nada$, in
order to avoid some bootstrapping issues in the definitions that
follow.

We can look at $A$ as a \emph{single stream} but we can also view $A$
as being comprised of \emph{several streams} at several
\emph{levels}. More precisely, for some \emph{minimal level} $m \in
\mb{N}$ we can define $A_m : \mb{R}^+ \to 2^{\mc{B}_{\ge m}}$ as
  the restriction of $A$ to $\mc{B}_{\ge m}$, i.e.: $A_m(t) =
  A(t) \cap \mc{B}_{\ge m}$. Since, by our current definition,
  the difficulty of mining a block at level $n$ is precisely twice as
  difficult as mining a block at level $n-1$, this means that we would
  expect the frequency of messages to go down at some point as we go
  up in the level of the message streams. 

When exactly the frequency will be low enough to establish consensus
will be determined by the total hashing power available in the network
as well as the time it takes a message to be become common
knowledge. In the remainder we will formulate the relevant criterion
that may be used to determine (from the perspective of a single peer)
when consensus has been reached.

So let $\Delta \in \mb{R}^+$ be our \emph{full-stop time duration
  parameter}. We say $A$ \emph{was quiet for level $n$ at time $t$}
iff no blocks of level $n$ or higher arrived in the interval
$(t-\Delta, t]$, i.e.: $\forall t' \in (t-\Delta, t] . A_n(t) =
  \nada$.

Based on this, we define a function $Q : \mb{R}^+ \to \mb{N}$ that
maps every time instance to the lowest level $n$ for which $A$ was
quiet at that time instance, i.e.: $Q(t) = \text{min}\{ n\ |\ A_n(t) =
\nada\}$.
 
%

We now use a transfinite recursion to define the \emph{ledger stream}
as a function $L: \mb{R}^+ \to 2^E$ mapping instances in time to the
maximal sets of entries that become part of the ledger at that time
(from the perspective of any of the peers). In particular we let $L(0)
= \nada$ and for $t > 0$ we let $e \in L(t)$ iff $e$ does not occur
earlier in the ledger stream yet at some preceding time $t' < t$ a
block $b \in A_{Q(t)}(t')$ arrived such that $e$ is part of $b$
(either directly or indirectly).

A peer can implement this consensus ledger by simply waiting for quiet
streams (on all levels) and, as soon a stream goes quiet, inserting
any and all so far unseen entries that are contained in the so far
untreated blocks in that stream. Please note that, although the
definition refers to \emph{all} blocks in the history of the stream,
in practice, once all entries that are part of a given block have
been entered into the ledger this block may be discarded from memory
(because it cannot affect the ledger anymore).

\section{Conclusion}

We have formalized a novel cooperative proof-of-work scheme, which
constitutes a refinement to the well known, and widely used,
proof-of-work scheme of \emph{zeroing a cryptographic hash}.
We have discussed some of the intuitions behind this cooperative
scheme as well as some of the potential benefits that the cooperative
scheme may offer over the non-cooperative scheme in practical
applications.
As future work we are interested in implementation and experimental
validation of this scheme.

\bibliography{main}

\begin{thebibliography}{1}

\bibitem{decker2013information}
Christian Decker and Roger Wattenhofer.
\newblock Information propagation in the bitcoin network.
\newblock In {\em Peer-to-Peer Computing (P2P), 2013 IEEE Thirteenth
  International Conference on}, pages 1--10. IEEE, 2013.

\bibitem{demers1987epidemic}
Alan Demers, Dan Greene, Carl Hauser, Wes Irish, John Larson, Scott Shenker,
  Howard Sturgis, Dan Swinehart, and Doug Terry.
\newblock Epidemic algorithms for replicated database maintenance.
\newblock In {\em Proceedings of the sixth annual ACM Symposium on Principles
  of distributed computing}, pages 1--12. ACM, 1987.

\bibitem{lewenberg2015bitcoin}
Yoad Lewenberg, Yoram Bachrach, Yonatan Sompolinsky, Aviv Zohar, and Jeffrey~S
  Rosenschein.
\newblock Bitcoin mining pools: A cooperative game theoretic analysis.
\newblock In {\em Proceedings of the 2015 International Conference on
  Autonomous Agents and Multiagent Systems}, pages 919--927. International
  Foundation for Autonomous Agents and Multiagent Systems, 2015.

\bibitem{zohar2015bitcoin}
Aviv Zohar.
\newblock Bitcoin: under the hood.
\newblock {\em Communications of the ACM}, 58(9):104--113, 2015.

\end{thebibliography}

\end{document}